\documentclass{aa}  
\usepackage{graphicx}
\usepackage{txfonts}
\begin{document}
   \title{The molecular environment of the massive star forming region NGC~2024: Multi CO transition analysis}
   \titlerunning{The Molecular Environment of NGC~2024}

   \subtitle{}

   \author{Emprechtinger M.
          \inst{1}
          \and Wiedner, M. C. \inst{1}
	  \and Simon, R. \inst{1}
	  \and Wieching, G. \inst{1,2}
	  \and Volgenau, N. H. \inst{1}\fnmsep\thanks{Current address: California Institute of Technology, Owens Valley
Radio Observatory, Big Pine, CA 93513   USA}
	  \and Bielau, F. \inst{1}
	  \and Graf, U.U. \inst{1}
	  \and G\"usten, R. \inst{2}
	  \and Honingh, C. E. \inst{1}
	  \and Jacobs, K. \inst{1}
	  \and Rabanus, D. \inst{1,3}
	  \and Stutzki, J. \inst{1}
	  \and Wyrowski, F. \inst{2}
          }
	  
   \offprints{M. Emprechtinger}

   \institute{I. Physikalisches Institut, Universit\"at zu K\"oln,
              Z\"ulpicher Str. 77, 50937 K\"oln, Germany
              \email{emprecht@ph1.uni-koeln.de}
         \and
             Max-Planck Institut f\"ur Radioastronomie, Auf dem H\"ugel 69, 53121 Bonn, Germany  
	  \and
	     European Southern Observatory, Alonso de Cordova 3107, Vitacura, Casilla 19001, Santiago, Chile}

   \date{}

 
  \abstract
   {Sites of massive star formation have complex internal structures. Local heating  by young stars and kinematic processes, such as outflows and stellar winds, generate large temperature and velocity gradients. Complex cloud structures lead to intricate emission line shapes.
   CO lines from high mass star forming regions are rarely Gaussian and show often multiple peaks. Furthermore, the line shapes vary significantly with the quantum number J$\rm _{up}$, due to the different probed physical conditions and opacities.}
   {The goal of this paper is to show that the complex line shapes of $^{12}$CO and $^{13}$CO in NGC~2024 showing multiple emission and absorption features, which vary with rotational quantum number $J$ can be explained consistently with a model, whose temperature and velocity structure are based on the well-established scenario of a PDR and the $''$Blister model$''$. }
   {We present velocity-resolved spectra of seven $^{12}$CO and $^{13}$CO lines ranging from J$\rm _{up}=3$ to J$\rm _{up}=13$. We combined these data with $^{12}$CO high-frequency data from the ISO satellite and analyzed the full set of CO lines using an escape probability code and a one-dimensional full radiative transfer code. }
   {We find that the bulk of the molecular cloud associated with NGC~2024 consists of warm (75~K) and dense ($\rm 9\cdot 10^5~cm^{-3}$)  gas. An additional hot ($\rm\sim 300~K$) component, located at the interface of the HII region and the molecular cloud, is needed to explain the emission of the high-J CO lines. Deep absorption notches indicate that very cold material ($\rm\sim 20~K$) exists in front of the warm material, too.}
   {A temperature and column density structure consistent with those predicted by PDR models, combined with the velocity structure of a $''$Blister model$''$,  appropriately describes the observed emission line profiles of this massive star forming region. This case study of NGC 2024 shows that, with physical insights into these complex regions and careful modeling, multi-line observations of $^{12}$CO and $^{13}$CO can be used to derive detailed physical conditions in massive star forming regions.
 }


   \maketitle
%
\section{Introduction}\label{intro}

High-mass star forming regions are very complex. 
The interaction of OB stars, which are often deeply embedded, with the surrounding molecular cloud via radiation and outflows and condensations of cold gas, still largely unaffected by the star forming activities, lead to a complex density and temperature structure, causing intricate shapes of the observed emission lines. Some CO lines in such regions (e.g., M~17, Stutzki~\&~Guesten et al.~\cite{m17}, W3, Kramer et al.~\cite{w3} and Mon~R2, Giannakopoulou et al.~\cite{monr2}) have multiple peaks, most likely due to absorption by foreground material. 
The line shapes vary significantly, depending on the rotational level and the observed isotopes, which trace different regimes of physical conditions and optical depths.  

The massive star forming region we selected for our study is the HII region NGC~2024 and its associated molecular cloud, which is located at a distance of 415~pc (Anthony-Twarog~\cite{tony}). It is a part of Orion~B, a well-studied giant molecular cloud (e.g. Maddalena et al.~\cite{mad}, Lada et al.~\cite{cs}, Kramer et al.~\cite{car}, Mitchell et al.~\cite{OB2}).

\begin{figure}[ht]
   {\resizebox{8cm}{!}
            {\includegraphics{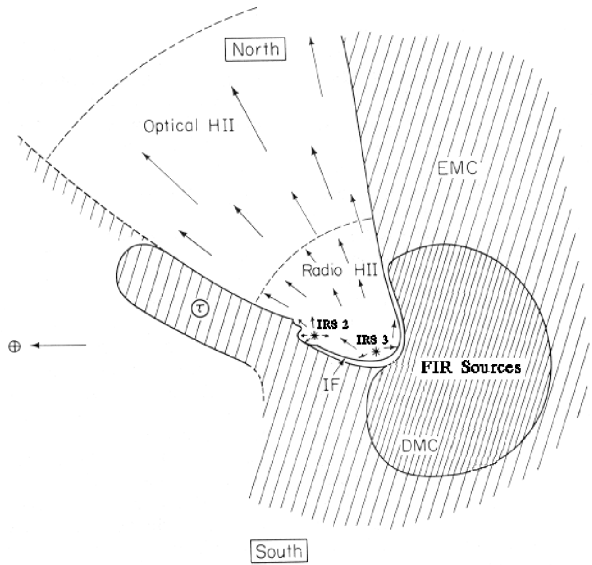}}}    
     \caption{Schematic view of the geometry of NGC2024. Abbreviations are as follows: DMC = dense molecular cloud, EMC = extended molecular cloud, IF=ionization front and $\tau$ = high optical depth cloud (optical dust bar). Adapted from Barnes et al.~(\cite{bar}) by Giannini et al.~(\cite{gia}).  }
         \label{geo}
  \end{figure}  

A possible scenario for the three-dimensional structure of the NGC~2024 region was proposed by Barnes et al.~(\cite{bar}), who combined observations from optical to radio wavelengths to construct a model of the cloud ( Fig.~\ref{geo}). In their model, the HII region sits in front of the bulk of the cloud but is partly obscured in the optical by a very prominent ridge of cold molecular material. This geometry provides an explanation for the complex line shapes, which may consist of contributions from fore- and background regions, including self-absorption.
The ionizing sources of NGC~2024 are invisible at optical wavelengths because they are obscured by the dust ridge. IRS2~b, a late O or early B star, has been identified as the primary ionizing source of NGC~2024 through observations in the near infrared (Bik et al.~\cite{bik}). The extinction ($\rm A_v$) through the dust ridge along the lines-of-sight to stars inside the HII region are in the range of 15 to 25$^{mag}$.
Furthermore, the HII region is expected to expand into the molecular cloud and to trigger star formation. Indeed some protostellar condensations (FIR 1 to 7) have been detected close to the HII region (Mezger et al.~\cite{1.3mm} \& \cite{mez}). Their masses, derived from sub-millimeter continuum emission, range from 1.6 to 5.1 M$_\odot$ (Visser et al.\cite{viss}). Because these masses correspond to visual extinctions between 270$^{mag}$ and 870$^{mag}$ and the optical depth of the dust ridge at the near side of the H~II region is only 15-25$^{mag}$, it is very likely that these condensations are located in the dense molecular cloud (DMC) behind the HII region. Massive outflows have been detected close to the source FIR~5 (e.g., Richer~\cite{out}, Sanders \& Willner~\cite{outflow}) giving additional evidence for ongoing star formation.

Detailed investigations of the molecular cloud associated with NGC~2024 were performed by Graf et al.~(\cite{urs1},~\cite{urs2}), who studied multiple $^{12}$CO lines (up to J=7-6) and its isotopologues. They derived temperatures of 23.5~K and 67.4~K for the foreground and the background component, as defined by Barnes et al.~(\cite{bar}), respectively. The corresponding column densities are $\rm 5.4\cdot 10^{22}~cm^{-2}$ (A$\rm _v=56~^{mag}$) and $\rm 2.0\cdot 10^{23}~cm^{-2}$ (A$\rm _v=210~^{mag}$)). Furthermore they found different velocities of 9~km/s and 11~km/s for the foreground and the background component, respectively. The peak velocities of the main component of the $^{12}$CO lines, v$\rm _{LSR}=13~km/s$, differ significantly from those of the optically thin lines (e.g., of C$^{18}$O), which are at a velocity of 11~km/s. Graf et al.~(\cite{urs2}) suggested that the blue side of the $^{12}$CO lines are absorbed by the dust ridge at a velocity of 9~km/s.   

This velocity shift between background and foreground component can be explained by a $''$Blister Model$''$ (Israel~\cite{blist}, Zuckerman~\cite{blist1}). In such a model, an OB star is assumed to be located close to the surface of the molecular cloud forming an HII region. The ionization front of such an HII region moves slowly into this cloud establishing a high pressure gradient. Because of this pressure gradient the ionized gas moves away from the molecular cloud. Therefore hydrogen recombination lines appear at  negative line of sight velocity offsets of about 3~km/s relative to the molecular lines, if the HII region is located at the near side of the molecular cloud (Israel~\cite{blist}). In the case of NGC~2024 the ionized material cannot flow into space, but pushes on the foreground material. We therefore expect that the foreground component is at somewhat lower velocities than the bulk of the molecular material. The assumption that the velocity structure of NGC~2024 is indeed caused by a $''$Blister$''$ is buttressed by the fact that the H109$\alpha$ recombination line appears at 7~km/s (Israel~\cite{blist}), and thus even more blueshifted.

C$^{18}$O~2-1 and C$^{17}$O~2-1 emission observed by Graf et al.~(\cite{urs2}), follows the 1.3~mm continuum very well and peaks close to the far infrared sources detected by Mezger et al.~(\cite{1.3mm} \& \cite{mez}). Contrary to the C$^{18}$O and C$^{17}$O lines, the map of the integrated intensity of optically thicker $^{12}$CO~7-6 line shows features similar to those seen in the 6~cm continuum map (Crutcher et al.~\cite{6cm}), which traces the ionized gas. This indicates that the $^{12}$CO~7-6 mainly originates from a photo dominated region (PDR) at the surface of the molecular cloud and not from embedded protostellar objects. The excitation conditions in this  PDR were studied using the integrated intensity of far infrared line emission observed with the ISO satellite (Giannini et al~\cite{gia}). They studied [NII], [NIII], and [OIII] lines from the HII region and high-J CO lines, [CII], and [OI], which originate (in part) from the PDR. Comparison with the predicted line ratios of a PDR model (Burton et al.~\cite{giapdr}) revealed a density of $\rm \approx 10^6~cm^{-3}$ and a UV field of $\rm 3\cdot 10^4~G_0$ at the surface of the molecular cloud.  

PDR models for high densities and high UV fields, as relevant for NGC~2024, predict  a column density for hot CO ($>100$~K) of the order of $\rm 10^{16}-10^{17}~cm^{-2}$ (see, R\"ollig et al.~\cite{R07}\footnote{All of their results are available under http://www.ph1.uni-koeln.de/pdr-comparison}). High-J CO lines (in this paper we  refer to all lines with frequencies $>1$~THz as high-J) would be emitted exclusively from such hot and dense material, because of their high critical density ($\rm \sim 10^7~cm^{-2}$) and high energy of the upper level (E$\rm _{up} \geq 250~K$).
 
In this paper we combine twelve transitions of $^{12}$CO and $^{13}$CO, ranging from J$\rm _{up}$=3 to J$\rm _{up}=19$ to trace molecular material NGC~2024 under very different physical conditions. The goal of this paper is to find, based on the above mentioned scenarios ($''$Blister Model$''$ and PDR) a model for the velocity, density and temperature structure of NGC~2024, which reproduces the observed line intensities and line shapes consistently. For this purpose we observed seven CO lines spectrally resolved which are presented in section~\ref{Obsres}. In section~\ref{Modres}, we first examine the physical conditions of the gas, which emits the high-J CO lines,  using an escape probability code and compare the result with the expectation from PDR models. The outcome of the escape probability calculations are used as input of a five component radiative transfer model (section~\ref{RTM}). An interpretation of the model is given in section~\ref{inter}.

\section{Observations}

\subsection{Observations of $\rm ^{12}CO~J=13-12$ at APEX}

The $^{12}$CO 13-12 ($\rm\nu=1496.922909~GHz$, M\"uller et al.~\cite{cdms}) observations were carried out on November~22,~2005 at the APEX 12 m telescope (G\"usten et al.~\cite{apex}), located on Llano de Chajnantor, Chile. We used the CO N$^+$ Deuterium Observations Receiver (CONDOR, Wiedner et al.~\cite{fl}), a heterodyne receiver that operates at THz frequencies (1.25-1.53~THz). The typical double sideband (DSB) receiver noise temperature was between 1500~K and 1900~K. The mean atmospheric transmission at zenith during the observations was $\sim~0.2$. As a backend, we used the APEX Fast Fourier Transform Spectrometer (FFTS), which has an intrinsic resolution of 60 kHz.

The expected diffraction limited main beam size of a 12~m telescope at 1.5~THz is $4.3''$. Because the accuracy of the primary surface derived from planet observations at frequencies between 350~GHz and 1500~GHz (with CONDOR) is 18~$\mu m$ (G\"usten et al.~\cite{apex}), we expect some of the power to be directed into an error beam. We assume that the error beam is approximately $80''$, which is derived from the approximate size of the individual panels of the telescope ($\sim 70$~cm). 
To determine the efficiency for the calibration to the main beam temperature (T$\rm _{MB}$), we measured the continuum of Moon and Mars and compared the observed values with models. These measurements lead to beam efficiencies of 0.4 and 0.09-0.11 for the Moon and Mars, respectively. The different beam efficiencies arise because Mars is about one-fourth the size of the error beam (18.2$''$ on the date of the observations), and thus these observations suffer from beam dilution.
A more detailed discussion of the beam sizes and coupling efficiencies is given in Volgenau et al.~(\cite{irc2}) and Wiedner et al.~(\cite{fl}).
 
Since NGC~2024 is an extended source, clearly larger than the error beam (Graf et al.~\cite{urs2}, Kramer et al.~\cite{car}), we use a beam efficiency of 0.4 for the calibration to T$\rm _{MB}$.  
Pointing and focusing were determined from observations of Mars, and we estimate a pointing accuracy better than $7''$. We observed $^{12}$CO~J=13-12 towards five positions, along a line from FIR~5 via IRS~3 to IRS~2. The coordinates of FIR~5 are $\rm\alpha = 5^h~41^m~44.18^s$, $\rm\delta = -1^\circ~55'~38.0''$ (J=2000, Mezger et al.~\cite{1.3mm}). The offsets of the other positions with respect to FIR~5 are listed in Tab.~\ref{pos}. 

\begin{table}[ht]
\caption{Positions of the $^{12}$CO~J=13-12 observations. The (0,0) position is $\rm\alpha = 5^h~41^m~44.18^s$, $\rm\delta = -1^\circ~55'~38.0''$ (J=2000, Mezger et al.~\cite{1.3mm}).} \label{pos}
\begin{center}
\begin{tabular}{lccc}
\hline
Position & $\Delta\alpha$ [$''$] & $\Delta\delta$ [$''$]  \\ 
\hline
\hline
FIR~5 & 0.0   & 0.0  \\
IRS~3 & 3.3   & 16.0 \\
\#1 & 8.8   & 28.6  \\
\#2 & 13.8  & 41.9  \\
IRS~2 & 24.6  & 68.5 \\
\hline
\end{tabular}
\end{center}
\end{table} 

\subsection{Observations of mid-J CO lines with KOSMA}

In addition to the $^{12}$CO~13-12 transitions, we used archival, so far unpublished, maps of $^{12}$CO~6-5 ($\nu =691.5$~GHz), $^{12}$CO~3-2 ($\nu =345.7$~GHz), $^{13}$CO~6-5 ($\nu =661.1$~GHz) and $^{13}$CO~3-2 ($\nu =330.6$~GHz). The observations were conducted between January 27 and February 2 1998 at the K\"olner Observatorium f\"ur Submillimeter Astronomie (KOSMA), located on Gornergrat, Switzerland. The beam sizes (HPBW) of these observations are $\sim 50''$ and $82''$ at 690~GHz and 345~GHz, respectively.  
To map the central part of the molecular cloud associated with NGC~2024, we used a dual-channel SIS receiver (Graf et al.~\cite{rec}) with typical DSB receiver noise temperatures of 100 and 400~K for the J=3-2 and J=6-5 lines, respectively.   
As backends, we used two acousto optical spectrometers (AOS). The Medium Resolution Spectrometer (MRS) has a bandwidth of 0.3~GHz and a resolution of 360~kHz. The Low Resolution Spectrometer (LRS) has a bandwidth of 1~GHz and a resolution of 1150~kHz. The forward efficiency, F$\rm _{eff}=0.93$, was determined by skydips.
For the J=3-2 observations, we used beam efficiencies (B$\rm _{eff}$) of 0.59 and 0.62 for the $^{12}$CO and the $^{13}$CO line, respectively. For the J=6-5 observations, the beam efficiencies were 0.40 and 0.48, respectively.

\subsection{Observations of mid-J CO lines with NANTEN2}

Simultaneous observations of $^{12}$CO~7-6 ($\nu = 806.65$~GHz) and $^{12}$CO~4-3 ($\nu = 461.04$~GHz) were carried out with the NANTEN2~4m telescope at Pampa La Bola, Chile. We obtained a $2' \times 2'$ map, which was centered on NGC~2024~IRS~3. The observations were carried out in December 2007, using the dual-channel 460/810~GHz test receiver, which had DSB receiver temperatures of $\sim 750$~K and $\sim 250$~K for the upper and the lower channel, respectively.  
Two AOS with bandwidths of 1~GHz and channel widths  of $\sim$ 560~kHz were used as backends. 
The beam sizes (HPBW) of the observations were $25''$ and $37''$ for the $^{12}$CO~7-6 and $^{12}$CO~4-3 observations, respectively. The beam efficiencies are 0.5 and 0.45 for 460 and 810~GHz, respectively, and a forward efficiency of 0.86 was measured at both frequencies.
The position of IRS~3 was taken to be $\rm\alpha = 5^h~41^m~44.40^s$, $\rm\delta = -1^\circ~55'~22.8''$ (J=2000, Mezger et al.~\cite{mez}). Pointing was checked on IRc2 in Orion A right before the observations and is expected to have an accuracy of $<7''$. 

\section{Observational Results}\label{Obsres}

To be able to compare the observed lines with each other, we convolved the maps observed with the KOSMA telescope and the NANTEN2 telescope to a spatial resolution of 80$''$, which is the spatial resolution of the J=3-2 spectra as well as the approximate size of the error beam of the $^{12}$CO~13-12 observations. Since 70\%-80\% of the radiation is expected to come from the error beam at these high frequencies, we considered 80$''$ as the spatial resolution of the $^{12}$CO~13-12 observations as well. An analysis of the maps of the lower-J lines will be given in a subsequent paper. 

\begin{figure*}[ht]
\begin{center}
   {\resizebox{16cm}{!}
            {\includegraphics{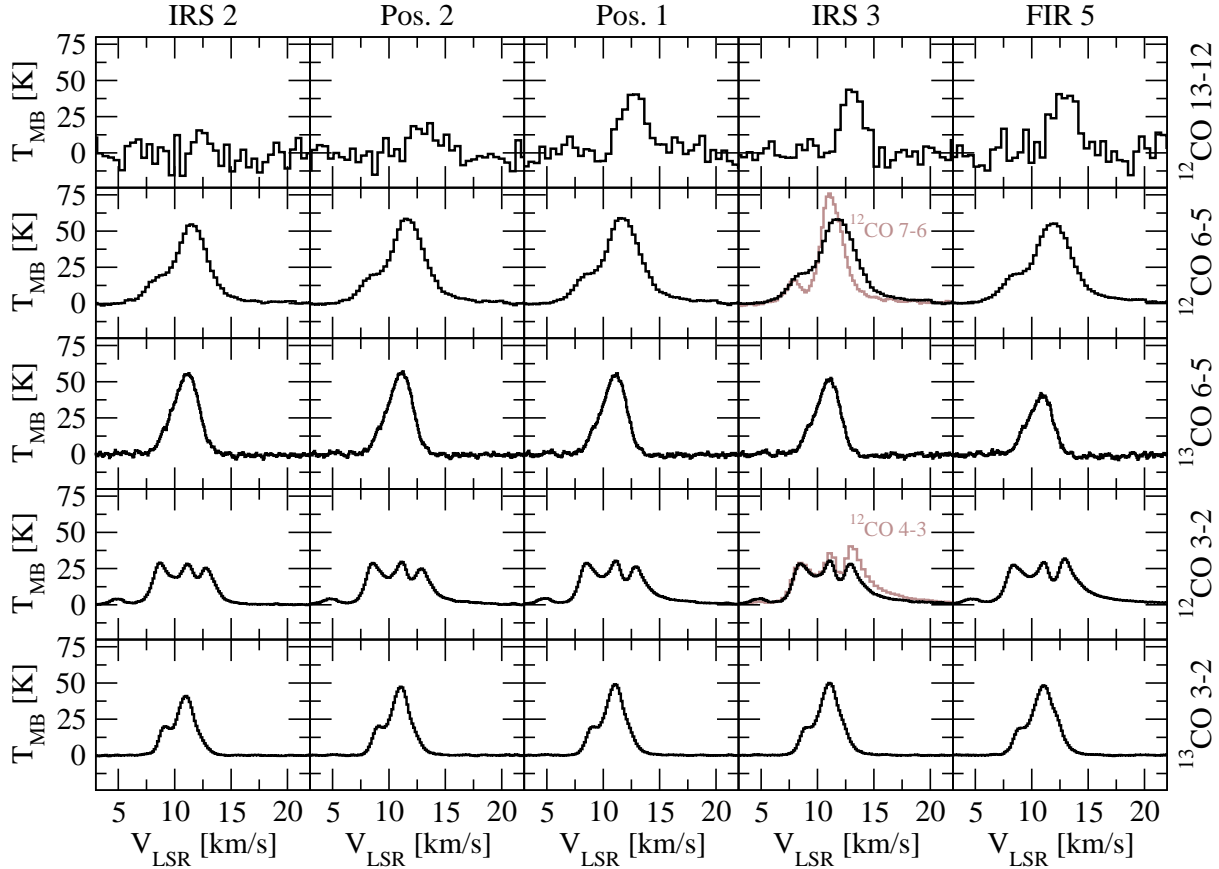}}}
\end{center}    
     \caption{Spectra of CO emission from pointed observations in NGC~2024. The spectra are all convolved to a resolution of $80''$, which is approximately  the resolution of the observed $^{12}$CO~3-2 and the $^{13}$CO~3-2 spectra as well as the resolution of the error beam of the $^{12}$CO~13-12 observations. The spectra are taken towards the positions (north to south left to right) as denoted in Tab.~\ref{pos}. The grey spectra, overlaid on $^{12}$CO~6-5 and $^{12}$CO~3-2 at ISR~3, are the $^{12}$CO~7-6 and the $^{12}$CO~4-3 spectra, respectively.}
         \label{spectra}
  \end{figure*}

The spectra are shown in Fig.~\ref{spectra} and the line parameters are listed in Tab.~\ref{linpara}. While most of the lines have a Gaussian line profile ($^{12}$CO~13-12, $^{13}$CO~6-5) or only  relatively weak blue shifted shoulders ($^{12}$CO~7-6, $^{12}$CO~6-5, $^{13}$CO~3-2), the $^{12}$CO 4-3 and 3-2 lines show complex line shapes with absorption notches, enhanced emission from the red shifted wing, and an additional bump at $\sim 4.5$~km/s. 

The emission of the J=3-2 and J=6-5 transitions of $^{12}$CO and $^{13}$CO are fairly uniform in line intensity and shape at all five positions, and the integrated intensity of these lines drops towards IRS~2 only by $\sim 30\%$. This uniformity is expected, since the separation between the positions of FIR~5 and IRS~2 is only $73''$, which is on the order of the spatial resolution of the observations. The $^{12}$CO~13-12 line peaks towards the southern three positions (FIR~5, IRS~3 and position \#1) and declines noticeably towards the north, indicating that high-J CO emission originates from a rather compact region around IRS~3.
  
Because of the different and complex line shapes of the $^{12}$CO and $^{13}$CO lines, we give the centroid velocity of the spectra in the following. The velocities of the $^{13}$CO lines are $\sim 11$~km/s, whereas the velocity of $^{12}$CO~3-2 and $^{12}$CO~6-5 is $\sim 12$~km/s. The J=13-12 line of $^{12}$CO is found at a velocity of $\sim 13$~km/s, although there is a variation in the velocity of about 1~km/s with position. The velocity of the $^{13}$CO lines is uniform throughout all five positions, whereas the velocity of $^{12}$CO~3-2 and $^{12}$CO~6-5 declines from south to north. This is due to a redshifted outflow detected in these two lines, which is strongest at the southern positions and reaches velocities up to $\sim 20$~km/s. This outflow was detected previously by Sanders \& Willner~(\cite{outflow}).
The lack of a blueshifted wing in our spectra is consistent with previous observations (e.g., Richer~\cite{out} \& Richer et al.~\cite{john2}). One explanation given by Richer et al.~(\cite{john3}) is that the outflow exists in a very complex region close to the ionization front of the HII region. North of its origin lies very dense gas seen in 1.3~mm map (Mezger et al.~\cite{mez}) and HCO$^+$ (Richer et al.~\cite{john3}). Any material ejected by the protostar in this direction might be retarded by dense gas and/or destroyed by strong UV-radiation. 



\begin{table}[ht]
\caption{Here the integrated intensities, v$\rm _{LSR}$ and $\Delta v$ of the lines shown in Fig.~\ref{spectra} are listed. Because of the complex line shapes, especially from the lower-J $^{12}$CO lines, the v$\rm _{LSR}$ and $\Delta v$ have been determined by using the first and second moment of the spectra, respectively. } \label{linpara}
\begin{center}
\begin{tabular}{lccc}
\hline
Position & $\rm \int T _{mb}~dv$ & V$\rm _{LSR}$ & $\Delta v$  \\ 
&  [Kkm/s] & [km/s] & [km/s] \\
\hline
\hline
\multicolumn{4}{c}{$\rm ^{12}CO~13-12$} \\
\hline
FIR~5 & $120\pm 10$ & $12.23\pm 0.3$ & $3.34\pm 0.5$\\
IRS~3 & $109\pm 10$ & $12.75\pm 0.3$ & $3.92\pm 0.5$\\
\# 1  & $127\pm 10$ & $13.50\pm 0.3$ & $4.51\pm 0.5$\\
\# 2  & $57\pm 10$  & $13.17\pm 0.3$ & $3.54\pm 0.5$ \\
IRS~2 & $<30$       & -- & -- \\
\hline
\hline
\multicolumn{4}{c}{$\rm ^{12}CO~7-6$} \\
\hline
IRS~3 & $229\pm 10$ & $11.5\pm 0.2$ & $ 4.5\pm 0.5$ \\
\hline
\hline
\multicolumn{4}{c}{$\rm ^{12}CO~6-5$} \\
\hline
FIR~5 & $288\pm 10$ & $12.0\pm 0.15$ & $7.5\pm 0.5$\\
IRS~3 & $289\pm 10$ & $11.9\pm 0.15$ & $7.2\pm 0.5$\\
\# 1  & $279\pm 10$ & $11.9\pm 0.15$ & $7.0\pm 0.5$\\
\# 2  & $265\pm 10$ & $11.9\pm 0.15$ & $7.2\pm 0.5$\\
IRS~2 & $235\pm 10$ & $11.8\pm 0.15$ & $7.5\pm 0.5$\\
\hline
\hline
\multicolumn{4}{c}{$\rm ^{13}CO~6-5$} \\
\hline
FIR~5 & $159\pm 5$ & $10.9\pm 0.15$ & $2.8\pm 0.5$  \\
IRS~3 & $157\pm 5$ & $10.9\pm 0.15$ & $2.7\pm 0.5$\\
\# 1  & $149\pm 5$ & $10.8\pm 0.15$ & $2.6\pm 0.5$\\
\# 2  & $138\pm 5$ & $10.7\pm 0.15$ & $2.5\pm 0.5$\\
IRS~2 & $106\pm 5$ & $10.5\pm 0.15$ & $2.3\pm 0.5$\\
\hline
\hline
\multicolumn{4}{c}{$\rm ^{12}CO~4-3$} \\
\hline
IRS~3 & $257.8\pm 10$ & $11.5\pm 0.2$ & $ 11.8\pm 0.3$ \\
\hline
\hline
\multicolumn{4}{c}{$\rm ^{12}CO~3-2$} \\
\hline
FIR~5 & $213\pm 10$ & $11.9\pm 0.15$ & $10.2\pm 0.2$\\
IRS~3 & $202\pm 10$ & $11.9\pm 0.15$ & $11.3\pm 0.2$\\
\# 1  & $186\pm 10$ & $11.6\pm 0.15$ & $10.7\pm 0.2$\\
\# 2  & $171\pm 10$ & $11.0\pm 0.15$ & $8.55\pm 0.2$\\
IRS~2 & $155\pm 10$ & $10.7\pm 0.15$ & $6.52\pm 0.2$\\
\hline
\hline
\multicolumn{4}{c}{$\rm ^{13}CO~3-2$} \\
\hline
FIR~5 & $146\pm 5$ & $10.9\pm 0.1$ & $3.36\pm 0.2$\\
IRS~3 & $144\pm 5$ & $10.8\pm 0.1$ & $3.42\pm 0.2$\\
\# 1  & $136\pm 5$ & $10.8\pm 0.1$ & $3.34\pm 0.2$\\
\# 2  & $127\pm 5$ & $10.8\pm 0.1$ & $3.24\pm 0.2$\\
IRS~2 & $109\pm 5$ & $10.7\pm 0.1$ & $2.93\pm 0.2$\\
\hline 

\end{tabular}
\end{center}
\end{table} 

The $^{12}$CO~13-12 data displayed in Fig.~\ref{spectra} are binned to a resolution of 0.48~km/s, which is sufficiently high, because the line width of the J=13-12 line is on the order of 2.5 km/s. At this resolution we detected the line in four out of five positions with an S/N ratio $>3\sigma$. At the position of IRS~2, we can claim only a tentative detection of $\rm 7.5\pm 3~K$. 

The $^{12}$CO~6-5 line is not symmetric, but shows a shoulder at its lower velocity side.
Its velocity of $\rm\sim 12~km/s$ is approximately 1~km/s redshifted with respect to the $^{13}$CO~6-5 line. The difference in velocity can be explained by the two-component model, suggested by previous studies (Barnes et al.~\cite{bar}, Graf et al.~\cite{urs2}). 
In this model, the foreground component ($\tau$ in Fig.~\ref{geo}) is assumed to be at a lower v$\rm _{LSR}$ than the background component (9.2~km/s and 11.1~km/s, respectively). Furthermore, the temperature of the foreground component (24~K) is  lower than the temperature    of 67~K of the background (Graf et al.~\cite{urs2}).
Thus, the foreground component adds an absorption feature centered at 9.2~km/s to the $^{12}$CO~6-5 line emitted from the background component. Therefore, the centroid velocity of the spectra appears redshifted with respect to the background component. In $^{13}$CO~6-5, no signs of self-absorption appear, and thus its centroid velocity coincides with the v$\rm _{LSR}$ of the background component. 

The profile of the $^{13}$CO~3-2 line looks similar to the \mbox{$^{12}$CO~6-5} line, but it does not appear redshifted and lies at a velocity of $\sim 11$~km/s. 
The shape of the $^{12}$CO~3-2 lines is dominated by deep absorption notches at velocities of 9.8~km/s and 11.9~km/s, which are caused by material located in front of the HII region. The alternative scenario that the emission is composed of three individual, equally strong ($\sim 25$~K), velocity shifted cloud components is unlikely, because these components are not seen in most of the other lines.
A fourth, weaker peak ($\rm\sim 3~K$) can be seen at 4.6~km/s. $^{12}$CO~2-1 observations (Kramer et al.~\cite{car}) revealed that this component extends further to the north-east and seems to be kinematically distinct from the main component.


The spectra of $^{12}$CO~7-6 and $^{12}$CO~4-3, convolved to a resolution of 80$''$, are superimposed on the $^{12}$CO~6-5 and $^{12}$CO~3-2 in Fig.~\ref{spectra}, respectively. The $^{12}$CO~4-3 spectrum looks very much like the one of $^{12}$CO~3-2. The two absorption notches, the 4.6~km/s component, and the redshifted outflow can be seen in both spectra. Even the intensities are nearly identical. The only difference is that the emission from the outflow, i.e., at v$\rm _{LSR}>12~km/s$, is slightly stronger in $^{12}$CO~4-3. 

The blueshifted shoulder of the $^{12}$CO~6-5 spectrum, interpreted as self absorption, appears similar but more pronounced in $^{12}$CO~7-6. However, the $^{12}$CO~7-6 line is narrower, because the outflow signal is weaker than in $^{12}$CO~6-5. $^{12}$CO~7-6 seems to suffer much more from absorption and what is seen as a shoulder  in $^{12}$CO~6-5 appears as a second peak here. 
The different strengths of the absorption of $^{12}$CO~7-6 and $^{12}$CO~6-5 requires a strong gradient in the excitation of those lines.
The velocity of the $^{12}$CO~7-6 (11.5~km/s) line is in between the velocity of $^{12}$CO~6-5 ($\sim 12$~km/s) and $^{13}$CO~6-5 ($\sim 11$~km/s). If these velocity differences are indeed due to foreground absorption, the velocity dispersion of the $^{12}$CO~7-6 absorbing material is significantly lower than the velocity dispersion of the material which absorbs $^{12}$CO~6-5. 
The minimum, most likely caused by the foreground absorption, lies at 9.5 km/s, which is 0.3 km/s blue-shifted with respect to the absorption notches seen in $^{12}$CO~4-3 and $^{12}$CO~3-2.  

Graf et al.~(\cite{urs2}) observed NGC~2024 in $^{12}$CO~7-6 as well, using the UKIRT telescope, which has a similar spatial resolution. Their spectrum at FIR~5, $16''$ south of IRS~3, shows a v$\rm _{LSR}$ of 12.8~km/s, clearly red shifted with respect to our observations. However the frequency stability of the Laser system that was used as local oscillator in these early measurements was about 1 MHz ($\sim 0.4$~km/s). Furthermore, the UKIRT observations were made in a double beam switch mode with a  chop throw of $3'$. Thus their off-position  might have been not completely clean, leading to a different apparent line shape. Possible pointing errors cannot be the reason for different velocities reported in Graf et al.~(\cite{urs2}) and this work, because no big velocity gradients with position are seen in NGC~2024.

\section{Modeling Results}\label{Modres}

The aim of this investigation is to show that the major feature of the complex shapes and intensities of all observed $^{12}$CO and $^{13}$CO lines can be explained consistently with a physically plausible scenario, which is based on the $''$Blister model$''$ and the PDR scenario (see section~\ref{intro}).

\subsection{Escape Probability Code Results}\label{epcs}

In this section we examine the physical properties of the high-J CO lines using an escape probability code (Stutzki \& Winnewisser~\cite{epc}). By comparing the column density of hot CO required to explain the observed high-J CO emission with the column density expected by PDR models we check, if the assumption of a PDR-scenario is consistent. Furthermore we use the results of this models as a first guess for a more sophisticated five component model (section~\ref{RTM}), which reduces the free parameter of this model. 

In addition to the new $^{12}$CO~13-12 spectra of NGC~2024 presented in this work, the integrated CO intensities from J$\rm _{up}$=14 to J$\rm _{up}$=17 observed with the ISO satellite towards FIR~5 (Giannini et al.~\cite{gia}) were taken from literature. The $^{12}$CO~13-12 line does not show any sign of absorption or emission at $\sim 9,5$~km/s. Thus, we assume that the emission of $^{12}$CO~13-12, as well as of the other high-J CO lines, is purely determined by the material located behind the HII region.  
Therefore, we can use an escape probability code (Stutzki \& Winnewisser~\cite{epc}) to model the emission of $^{12}$CO $\rm J\geq 13$.
In this code, the emitting gas is treated as an isothermal cloud with homogeneous density. For a given set of kinetic temperatures (T$\rm _{kin}$), H$_2$ densities (n(H$_2$)), and $^{12}$CO column densities per velocity interval (N(CO)/$\Delta$v), the emitted line strengths (main beam temperatures and integrated intensities), as well as the optical depths ($\tau$) at the line center are calculated. 


To match the observed intensities of the high-J lines, a $^{12}$CO column density of $\rm 9.5\pm 0.7\cdot 10^{16}~cm^{-2}$ in a velocity interval $\Delta v=4$~km/s, i.e., the width of the $^{12}$CO~13-12 line, is required. This column density is in good agreement with PDR models (see section~\ref{intro}). Constraints on the gas temperature and H$_2$-density, however, are looser; many solutions are possible. The two best fit solutions (both with a $\chi ^2$ of 2.9) yield a  H$_2$-density of $\rm 1.3\cdot 10^6~cm^{-3}$ and a temperature of 250~K and $\rm n_{H_2}=3\cdot 10^5~cm^{-3}$ and T=410~K, respectively. All models with a $\chi ^2$ lower than 10 have in common that  $\rm T^3\cdot n$ is approximately constant.


If we assume that the gas is distributed uniformly throughout the ($80''$) beam (and the area filling factor is one), then this layer of the PDR has a thickness of  $5\cdot 10^{14}$ to $1.5\cdot 10^{15}$~cm (= 33-100~AU). At a distance of 415~pc, 100~AU corresponds to $0.25''$, which indicates that this hot component is indeed a thin layer at the, possibly clumpy, surface of the molecular cloud. 



\subsection{Full Radiative Transfer Model}\label{RTM}

\begin{table*}[ht]
\begin{center}
\caption{Physical properties of a 5-component radiative transfer model of NGC~2024. Properties with a range of values vary continuously with depth.} \label{simres}
\begin{tabular}{ccccccc}
\hline
Component & T [K] & n(H$_2$) [cm$^{-3}$] & r [cm] & N(H$_2$) [cm$^{-2}$] & v$\rm _{LSR} [km/s]$ & $\rm\Delta v_{turb}$ [km/s] \\ 
\hline
\hline
B1 & 75 & $9\cdot 10^5$ & $8.0\cdot 10^{16}$ &$ 7.2\cdot 10^{22}$ & 11.0 & 1.8 \\
B2 & 75-330 & $9\cdot 10^5$-$2\cdot 10^6$ & $6.2\cdot 10^{13}$ & $8.9\cdot 10^{19}$ & 11.0 & 1.8 \\
F1 & 330-40 & $ 1\cdot 10^5$ & $8.9\cdot 10^{14}$ & $8.9\cdot 10^{19}$ & 9.3 & 1.3 \\
F2 & 40-30 & $ 1\cdot 10^5-3\cdot 10^4 $ & $1.7\cdot 10^{17}$ &$ 1.04\cdot 10^{22}$ & 9.3-9.7 & 1.3-1.1 \\
F3 & 30-15 & $ 5\cdot 10^4 $ & $5.9\cdot 10^{15}$ & $ 2.9\cdot 10^{20}$ & 12.1 & 1.8 \\
\hline
\end{tabular}
\end{center}
\end{table*} 

To get a more complete picture of the source, which explains also the complex line shapes of the lower-J lines we have to use a multilayer radiative transfer model, which treats multiple emission components and self-absorption correctly. With such a radiative transfer model, we fitted the line shapes and intensities of all seven $^{12}$CO and $^{13}$CO lines observed at the position of IRS~3 and the integrated intensities of $^{12}$CO with J$\rm _{up}>13$, simultaneously. The $^{12}$CO line with J$\rm _{up}>13$ are the same data we used in the previous section. These data have been observed at the position of FIR~5, which lies 16$''$ south of IRS~3, a separation much lower than the spatial resolution of our observations.
We used SimLine (Ossenkopf et al.~\cite{sim}), a 1-dimensional, spherical radiative transfer code. SimLine computes the  profiles of molecular rotational lines for an arbitrary density, temperature and velocity distribution, specified as a set of discrete layers, by integrating the radiative transfer equation numerically. The population of the individual levels of a molecule are computed by solving the full system of balance equations iteratively. By setting the inner radius of the model to a value, which is much larger than the beamsize times the distance of the observations, we used SimLine in a quasi plane-parallel way.

\subsubsection{Constraints on the models} 
 
A close look at the data in Fig.~\ref{spectra} and Table~\ref{linpara} reveals an apparent contradiction in the line strengths at v$\rm _{LSR}=11$~km/s. The strong ($\sim 50$~K) $^{13}$CO~6-5 emission, which indicates a large column density of hot molecular gas, belies the relatively weak ($<15$~K) $^{12}$CO~13-12 emission at the same velocity.
The possibility that the $^{12}$CO~13-12 line is absorbed by the foreground gas can be ruled out, since the estimated column density ($\rm N_{H_2}=2\cdot 10^{22}~cm^{-2}$, which corresponds to A$\rm _v$=20-25$^{mag}$, Bik et al.~\cite{bik}) is orders of magnitudes too low for the derived temperatures ($\sim 25$~K) to cause this absorption (Graf et al.~\cite{urs2}). 
Relatively strong $^{12}$CO~13-12 emission at a v$\rm _{LSR}$ of $\sim 13$~km/s, where no other optically thin line shows an emission peak, suggests an error in the frequency calibration. The origin of such an error is puzzling, given that particular care was taken to check calibration during the observations. Immediately prior to observing NGC~2024, we observed Orion FIR~4 to check telescope pointing, and the $^{12}$CO~13-12 emission had the expected velocity   (Wiedner et al.~\cite{fl}, Kawamura~\cite{kaw}, Wilson et al.~\cite{wil}). The uniformity of the v$\rm _{LSR}$ throughout the $^{13}$CO~6-5 map also diminishes the likelihood that the anomalous $^{12}$CO~13-12 velocity is due to a pointing error. 
Moreover, all software-set values (e.g., rest frequency and sky frequency of the line) were checked carefully, but no  error was found.
Despite this scrutiny we are forced to assume that the $\rm v_{LSR}$ of $^{12}$CO~13-12 and $^{13}$CO~6-5 are both 10.9~km/s, to get a reasonable solution for our model. 
This assumption is buttressed by unpublished high-J CO observations (Marrone, priv. comm.), which also indicate a v$\rm _{LSR}$ of $\sim 11$~km/s.   

In addition to the previous two component models we introduced a hot, dense, and thin gas component, located at the interface of the HII region and the DMC (Fig.~\ref{geo}). An independent estimation of the physical conditions of this hot material, which is illuminated by OB stars within the HII region, has been derived with the escape probability code. 
In the case of the foreground material, we have to assume several components as well. First, we have to adopt two components of cold gas at different velocities, as both the $^{12}$CO~4-3 and the $^{12}$CO~3-2 spectra show two absorption notches. In addition, it is reasonable that material located at the interface of the foreground component and the ionized gas is heated as well, and thus we introduced  a layer of hot gas on this side of the HII region, too. Furthermore, the total column density of all components located in front of the HII region is restricted by the measurements of A$\rm v$ towards the obscured stars, which translates into a total H$_2$ column density of $\rm 2\cdot 10^{22}~cm^{-2}$.



\subsubsection{Best Fit Model: Reproducing the observed line profiles}  

The result of the modeling is listed in Tab.~\ref{simres}. As mentioned above, five components (three foreground and two background components) were appropriate to get a reasonable solution for the observed spectra. 
Each these five components is described by 5 parameters ({\it{T, n(H$_2$), r, v$_{LSR}$ and $\Delta v$}}) and thus the model contains 25 parameters. However, since the hot foreground component (F1) is determined by the temperature and column density of the hot background component (B2) and the density, v$\rm _{LSR}$ and $\Delta$v of the bulk of the foreground material (F2), these five parameters are not free. In addition, we set  v$\rm _{LSR}$ and $\Delta$v of the background components (B1 and B2) to the same values, and thus the number of free parameters is 18. But even these 18 parameter are not fully free. The total column density of the foreground components (F1+F2+F3)) must correspond to an A$\rm _v$ of 10-25$^{mag}$ to match the IR-observation, the temperature and the column density of the hot background component (B2) has to be consistent with the prediction of the PDR models, and the velocity difference between the background components and the bulk of the foreground material must be negative and in the range of a few km/s in order to be explained by the blister model. Therefore the number of effectively free parameters is $\sim 15$.

The background material was divided into two subcomponents, denoted by B1 and B2. B1 consists of warm (T$= 75$~K) and  dense gas (n(H$_2$)$= 9\cdot 10^5$~cm$^{-3}$) with a high column density (N$\rm (H_2) = 7.2\cdot 10^{22}~cm^{-2}$). B2 is a thin layer with a steep temperature and density gradient. At the surface of the cloud the temperature reaches 330~K and the density is as high as $\rm 2\cdot 10^6~cm^{-3}$. Both B1 and B2 have a velocity of v$\rm _{LSR}=11$~km/s and a FWHM of the turbulent velocity of $\rm \Delta v_{turb}=1.8~km/s$.

The foreground component was divided into three subcomponents, F1, F2, and F3, to explain the two absorption dips in the $^{12}$CO~4-3 and $^{12}$CO~3-2 spectra and for physical reasons, as mentioned above. The v$\rm _{LSR}$ of F2 shows a gradient from 9.3~km/s to 9.7~km/s towards the observer, and the v$\rm _{LSR}$ of F3 is 12~km/s. 
The total column density of the foreground component (F1+F2+F3; N$\rm (H_2)_{tot}=1.08\cdot 10^{22}~cm^{-2}$) correspond to an optical extinction of $\rm \sim 11^{mag}$, which is slightly lower than the 15 to 25$\rm ^{mag}$ found by Bik et al.~(\cite{bik}). To convert the column density into A$\rm _v$ we used the relation $\rm N_{H_2}[cm^{-2}]=0.94\times10^{21}A_v~(mag)$. However, since the light of some of these stars might be extincted by dust in the vicinity of the star, the lower value of A$\rm _v$ is probably more appropriate. 

\begin{figure}[ht]
   {\resizebox{8.7cm}{!}
            {\includegraphics{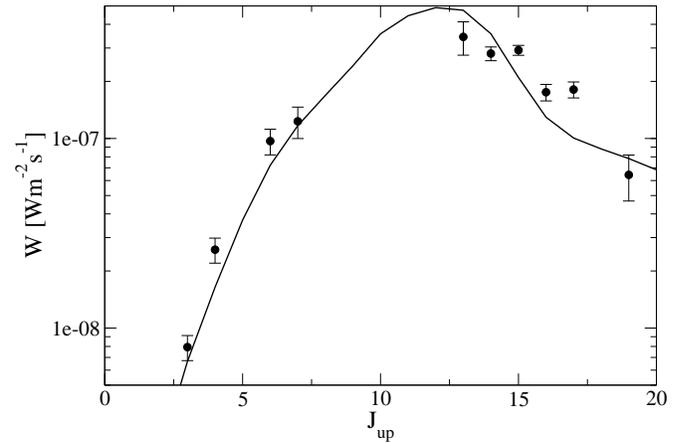}}}    
     \caption{Integrated line intensity of $^{12}$CO versus J$\rm _{up}$. The observed data (full circles) include all $^{12}$CO lines from this work and observations with the ISO satellite from Giannini et al.~(\cite{gia}). The solid line shows the results of the full radiative transfer code SimLine}
         \label{FJup}
  \end{figure}

The complex shapes of the emission lines, especially the deep absorption notches in $^{12}$CO~3-2, require depth-dependent gradients in the foreground subcomponents. F2 contains most of the foreground material ($> 96~\%$ of the mass). It consists of a $1.54\cdot 10^{17}$~cm deep, 40~K warm gas layer, whose density decreases linearly from $10^5$~cm$^{-3}$ to $3\cdot 10^4$~cm$^{-3}$. Subsequently the F2 gas density stays constant, but the temperature drops to 30~K within the next $1.4\cdot 10^{16}$~cm. Simultaneously to the temperature decrease, the v$\rm _{LSR}$ shifts from 9.3~km/s to 9.7~km/s. F3 is a comparatively thin ($\rm N_{H_2}=2.9\cdot 10^{20}~cm^{-2}$) component with a constant density of $5\cdot 10^{4}$~cm$^{-3}$and a temperature that drops from 30~K to 15~K. The v$\rm _{LSR}$ of F3 is 12.2~km/s. F3 is responsible for one of the two absorption dips in $^{12}$CO~3-2, as well as the absorption of the $^{12}$CO~6-5 intensity. 

The F1 component represents the radiation heated counterpart of B2 on the nearside of the HII region. This additional gas component does not change the modeled line emission as long as the H$_2$ column density is lower than $\rm 4.6\cdot 10^{20}~cm^{-2}$, i.e. as long as its column density is smaller than 5 times the column density of B2.  Because the existence of such a component is physically reasonable, we assumed that the interface between the bulk of the foreground material (F2) and the HII region is heated similar to the interface at the background, i.e., column density and maximum temperature of F1 and B2 are similar. Density and velocity of F1 are the same as found for F2.

\begin{figure}[ht]
   {\resizebox{8.7cm}{!}
            {\includegraphics{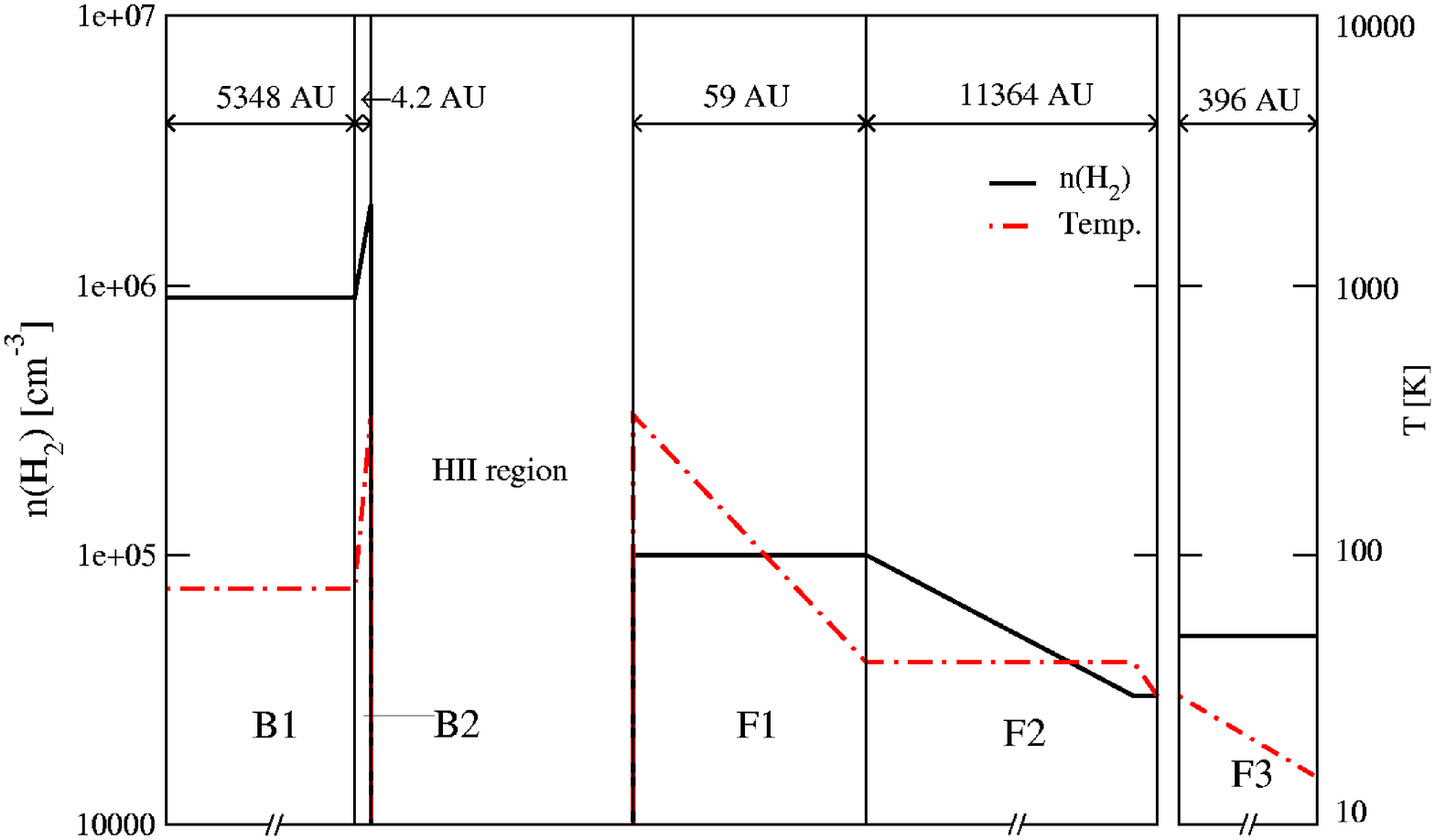}}}    
     \caption{Temperature and density structure of NGC~2024. The observer is located to the right. The component F3 resides on the near side of the HII region, but its exact position is not clear and it might be physically disconnected from the other components.}
         \label{scratch}
  \end{figure}

The geometric arrangement of the components B1, B2, F1, and F2 is well defined by our modeling and the temperature- and density trend of these components are shown in Fig.~\ref{scratch}. However, the location of F3 is not clear, apart from residing in front of the HII region. The fact that this component is cold and not blueshifted with respect to the background  component indicates that this component might be closest to the observer.

\begin{figure}[ht]
   {\resizebox{8.7cm}{!}
            {\includegraphics{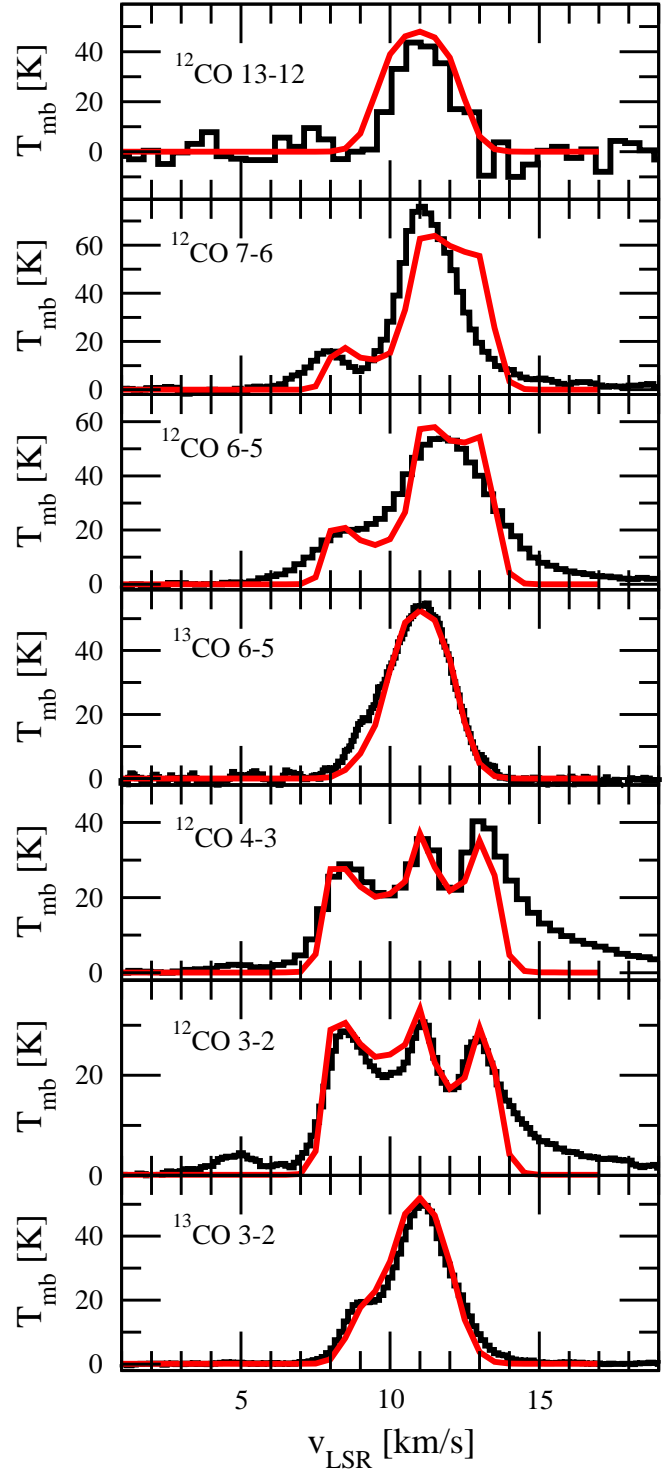}}}    
     \caption{Observed (black) and model (red) spectra of $^{12}$CO~13-12, $^{12}$CO~6-5, $^{13}$CO~6-5, $^{12}$CO~3-2 and $^{13}$CO~3-2. The observed spectrum of the $^{12}$CO~13-12 line was shifted by -2.2~km/s (see text).}
         \label{sline}
  \end{figure}

A comparison between the observed spectra and the model results is shown in  Fig.~\ref{sline}.
Given the known complexity of the source and the limits of a five-component model, the model spectra fit the observations quite well. Two spectral features that were not included in the model fit are the redshifted outflow, which can be seen in $^{12}$CO~6-5, $^{12}$CO~4-3 and $^{12}$CO~3-2, and the component at $\rm v_{LSR} \sim 4.5~km/s$. Nevertheless, the main features of the five emission lines can be explained with our rather simple assumptions. 

In Fig.~\ref{FJup} we show the integrated intensity of the $^{12}$CO lines versus J$\rm _{up}$. Especially for the low- and mid-J CO lines the model matches the observations quite well. At $\rm J_{up}>10$ the scatter of the observed data increases, possibly due to observational uncertainties.

\section{Physical interpretation and discussion of the model results}\label{inter}

The model discussed above, successfully explaining the observed lines including the profiles of the velocity resolved lines, is the simplest model scenario satisfying the constraints imposed by the blister and PDR picture.  

The B1 component is equivalent to the DMC in Fig.~\ref{geo}. Because the properties of B1 are mostly based on the fits of lines with J$\rm _{up}<5$, only the warm gas of the DMC is represented. Cold gas, which is deeper inside of the DMC, would have no effect on the modeled lines, either because the upper levels are not populated ( $^{12}$CO~13-12, $^{13}$CO~6-5), or the lines are already optically thick ($^{12}$CO~7-6 and $^{12}$CO~6-5). Thus, the column density of B1 is just the lower limit for the DMC. Observations of C$^{18}$O~2-1 and C$^{17}$O~2-1 (Graf et al.~\cite{urs2}) revealed a three times lager H$_2$ column density than what we found here. 
The B2 component represents the thin, hot interface between the HII-region and the molecular gas. Due to the high UV radiation emitted by the ionizing source(s) of NGC~2024, the gas at the interface is strongly heated, which may result in temperatures up to 330~K. 

The three subcomponents F1, F2, and F3 together represent the foreground material, $\tau$ in Fig.~\ref{geo}.
The F1 subcomponent is, as mentioned above, not traced by the lines we observed here, but it represents the counterpart of B2 at the foreground component albeit at lower density. For the other two subcomponents located in front of the HII region, F2 and F3, we assume that matter is distributed uniformly across the area of the beam. 
Interstellar matter, however, shows clumpy structure on all scales. In NGC~2024, substructures on scales down to the resolution of previous observations were detected (Kramer et al.~\cite{car}, Lada et al.~\cite{cs}, see Section~\ref{intro}). 
Therefore, we expect a clumpy structure within the beam as well, which would cause a much more complex structure of the foreground. Such small scale structures must be taken into account for more detailed modeling. For example, dense, cold clumps, smaller than the FWHM of the beam, which mainly affect the $^{12}$CO~3-2 line, but due to their low temperature leave the $^{12}$CO~6-5 line rather unaffected, are a possible explanation for the narrow absorption notches observed in $^{12}$CO~3-2 and 4-3, whereas the line shape of $^{12}$CO~6-5 is much smoother. This might be especially true for the  thin and cold component F3.  However, observations at higher spatial resolution would be necessary to verify this hypothesis.   

The velocity difference between the background gas and the bulk of the foreground material is 1.5~km/s. Assuming that the velocity difference is caused by the pressure of the ionized gas, according to the $''$Blister model$''$ this velocity shift corresponds to a  total momentum per unit area of $\rm 5670~g~cm^{-1}~s^{-1}$, which has to be transferred to the F2 component. We calculated the internal pressure of the ionized gas, which pushes the foreground component, using measured electron temperature (8160~K) and electron density of $\rm\sim 10^3~cm^{-3}$ (Odegard~\cite{ntd}). Such a pressure has to act for approximately $2\cdot 10^5$ years to accelerate the foreground component to its current velocity.
However, the mass of the F1 component is highly unknown, and T$_e$ and n$_e$ may have changed in the past, so that this simple calculation may be somewhat debatable, but it shows that the observed velocity structure can be indeed explained by a blister model. 


\subsection{Comparison with other models}  

The properties of the background components agree fairly well with the results obtained by Graf et al.~(\cite{urs2}). 
The main difference between their model and the present one is that Graf et al.~(\cite{urs2}) 
assumed LTE conditions, whereas in the SimLine code, the balance equations for all level populations and energy densities are solved self-consistently (Ossenkopf et al.~\cite{sim}).
In the previous study, an H$_2$ column density of the background component of $2.0\cdot 10^{23}$~cm$^{-3}$ was found, which is three times more than what we found here. This difference arises because they included rarer isotopes, e.g., C$^{18}$O and C$^{17}$O, which are optically thin and trace material located deep inside the cloud. Most of this material is hidden due to the high optical depths in the $^{12}$CO and $^{13}$CO lines we observed. The temperature determined by Graf et al.~(\cite{urs2}), 67.4~K, is in good agreement with the 75~K determined in our work. Also the kinematic parameters Graf et al.~(\cite{urs2}) report  (v$\rm _{LSR}=11.1~km/s$ and $\rm \Delta v=1.8~km/s$) match our results. Since they do not observe high rotational transitions of CO (no lines  with $\rm J_{up}> 7$), they have no information about the hot interface region. 

Giannini et al.~(\cite{gia}) give a CO column density of  $2-5\cdot 10^{18}$~cm$^{-2}$, which corresponds to an H$_2$ column density of $2.1-5.3\cdot 10^{22}$~cm$^{-2}$. This column density, which is about 50\% lower than the total column density we found ($\rm N_{H_2, total}=8.3\cdot 10^{22}~cm^{-2}$),  was derived using an escape probability code. The fact that Giannini et al. fitted only a single, isothermal gas component might explain the different results. In particular the background material will be underestimated in such an approach due to self absorption effects. 
The kinetic temperature they give (110-130~K) is in the range of the values we found, but because of their isothermal approach hard to compare with our results.


For the foreground components, we found a total H$_2$ column density of $1.07\cdot 10^{22}$~cm$^{-2}$, which corresponds to an A$\rm _v$ of 11.4. This is lower than the visual extinction of 15 to 25 found towards several stellar objects inside the HII region (Bik et al.~\cite{bik}).
However, due to the low spatial resolution of our observations, the beam-averaged column density might be well lower than the values obtained from individual stars.
The H$_2$ column density of the foreground component given by Graf et al.~(\cite{urs2}) is a factor of two larger than our value. They also found a temperature of 23.5~K, i.e., slightly lower than our results (30-40~K), which might be caused by the fact that they assumed LTE conditions.

\section{Conclusion}

We present the observations of seven $^{12}$CO and $^{13}$CO lines from J$\rm _{up}=3$ to J$\rm _{up}=13$ towards NGC~2024, a well know massive star forming region in Orion. The shapes of these lines range from almost Gaussian ($^{13}$CO~6-5, $^{12}$CO~13-12) to highly complex, multiple peaked ($^{12}$CO~3-2 , $^{12}$CO~4-3), which indicates a complex internal structure of NGC~2024. In our analyses we also included the integrated intensities of $^{12}$CO lines with J$\rm _{up}\geq 14$ (Giannini et al.~\cite{gia}).

We modeled the high-J CO lines, which all seem to be optically thin, using an escape probability code (Stutzki \& Winnewisser~\cite{epc}) and found that these lines are emitted from a hot ($\rm >250~K$), dense ($\rm >3\cdot 10^5~cm^{-3}$), and thin ($\sim 100$~AU) layer, which is most likely located at the interface of the HII region and the molecular cloud. 

We constructed a model for NGC~2024 using  1D radiative transfer code SimLine (Ossenkopf et al.~\cite{sim}). The velocity structure of this model is based on the principles of the Blister model and the temperatures and column densities of the components are constrained by the PDR scenario. This model explains the profiles of the observed $^{12}$CO and $^{13}$CO lines quite well. 
In our model, the bulk of the molecular gas resides at the back of the HII region and consists of warm (75~K) and dense ($\rm 9\cdot 10^5~cm^{-3}$) material. We also find  evidence for a hot (up to 330~K) and thin (400~AU) layer located at the surface of the molecular cloud, from which most high-J CO emission originates.    
The molecular ridge in front of the HII region consists of molecular material at lower densities ($\rm 3\cdot 10^4~cm^{-3}-10^5~cm^{-3}$). The column density is of the order of $\rm 10^{22}~cm^{-2}$, which correspond to an A$\rm _v$ of 11~mag. This value is in good agreement with measurements of the optical extinction towards stars within the HII region (Bik et al.~\cite{bik}). 

Overall, this study shows that for the example of NGC~2024, emission lines with complex and varying line shapes, as often observed in massive star forming regions, can be explained consistently with a rather simple, yet physically reasonable model.
To explain the  twelve emission lines a model of $\sim 15$ free parameter is required. Both the relatively large number of successfully fitted line intensities and line profiles and the fact that the multi-layer parameters are consistent with the physical scenario of a blister and a PDR, lead us to the conclusion that we found a plausible model for the warm molecular gas in NGC~2024.

Complex line profiles of low-J CO lines are commonly observed in massive star forming regions and with modern observatories (e.g., SOFIA, APEX and Herschel), more and more mid- and high-J CO observations will be available. This case study of NGC~2024 shows that with physical insight into these complex regions and careful modeling, the complex line-shapes and multi-line observations  can be used to derive valuable information on the physical conditions in massive star forming regions.


\begin{acknowledgement}

This publication is based on data acquired with the Atacama
Pathfinder Experiment (APEX). APEX is a collaboration between
the Max-Planck-Institut für Radioastronomie, the European Southern
Observatory, and the Onsala Space Observatory.

The KOSMA 3m observatory is administrated by the International Foundation Gornergrat \& Jungfraujoch. The universities of Cologne and Bonn are operating jointly the KOSMA~3m telescope and, together with the university of Nagoya, the NANTEN2 observatory. NANTEN2 is financially supported in part by a Grant-in-Aid for Scientific Research from the Ministry of Education, Culture, Sports, Science and
Technology of Japan (No. 15071203) and from JSPS (No. 14102003 and No.
18684003), and by the JSPS core-to-core program (No. 17004) and by special funding from the Land NRW.

The CONDOR receiver was built by the Nachwuchsgruppe of the
Sonderforschungsbereich 494, which is funded by the Deutsche Forschungsgemeinschaft
(DFG).

\end{acknowledgement}

\end{document}